\documentclass[10pt, conference, compsocconf]{IEEEtran}
\IEEEoverridecommandlockouts
\let\labelindent\relax
\usepackage{cite}
\usepackage{amsmath,amssymb,amsfonts}
\usepackage{algorithmic}
\usepackage{graphicx}
\usepackage{textcomp}
\usepackage{xcolor}

\usepackage{times}
\usepackage{helvet}
\usepackage{courier}
\usepackage{enumitem}
\usepackage{multirow}

\usepackage{amssymb}
\usepackage{times}
\usepackage{epsfig}
\usepackage{amsmath}
\usepackage{amssymb}
\usepackage{color}
\usepackage{stmaryrd}
\usepackage{adjustbox} 
\usepackage{bm}
\usepackage{balance}
\usepackage{flushend}
\usepackage{booktabs}

\usepackage{algorithm,algorithmic}

\makeatletter
\newcommand{\ALOOP}[1]{\ALC@it\algorithmicloop\ #1%
  \begin{ALC@loop}}
\newcommand{\ENDALOOP}{\end{ALC@loop}\ALC@it\algorithmicendloop}

\makeatother

\usepackage{booktabs} 
\usepackage{enumitem}
\usepackage{siunitx}
\usepackage{subfig}
\usepackage{tabularx}

\newtheorem{definition}{Definition} 

\newcommand{\sun}{\textsc{MEGA}}
\newcommand{\sunl}{\textsc{MEGA++}}

\def\BibTeX{{\rm B\kern-.05em{\sc i\kern-.025em b}\kern-.08em
    T\kern-.1667em\lower.7ex\hbox{E}\kern-.125emX}}
\begin{document}

\title{Joint Embedding of Meta-Path and Meta-Graph for \\Heterogeneous Information Networks}

\author{\IEEEauthorblockN{Lichao Sun\IEEEauthorrefmark{2},
Lifang He\IEEEauthorrefmark{3}\IEEEauthorrefmark{1}\thanks{\IEEEauthorrefmark{1}Corresponding author.},
Zhipeng Huang\IEEEauthorrefmark{4},
Bokai Cao\IEEEauthorrefmark{5},
Congying Xia\IEEEauthorrefmark{2},
Xiaokai Wei\IEEEauthorrefmark{5} and
Philip S. Yu\IEEEauthorrefmark{2}}
\IEEEauthorblockA{\IEEEauthorrefmark{2}University of Illinois at Chicago, Chicago, IL \IEEEauthorrefmark{3}Cornell University, New York City, NY }
\IEEEauthorblockA{\IEEEauthorrefmark{4}The University of Hong Kong, China \IEEEauthorrefmark{5}Facebook, Menlo Park, CA}
\IEEEauthorblockA{
Email: \{lsun29, cxia8, psyu\}@uic.edu, \{lifanghescut,caobokai.why,weixiaokai\}@gmail.com, zphuang@cs.hku.hk}
}

\maketitle

\begin{abstract}
Meta-graph is currently the most powerful tool for similarity search on heterogeneous information networks, where a meta-graph is a composition of meta-paths that captures the complex structural information. 
However, current relevance computing based on meta-graph only considers the complex structural information, but ignores its embedded meta-paths information. 
To address this problem, we propose MEta-GrAph-based network embedding models,
called {\sun} and {\sunl}, respectively.
The {\sun} model uses normalized relevance or similarity measures that are derived from a meta-graph and its embedded meta-paths between nodes simultaneously,
and then leverages tensor decomposition method to perform node embedding.
The {\sunl} further facilitates the use of coupled tensor-matrix decomposition method to obtain a joint embedding for nodes, which simultaneously considers 
the hidden relations of all meta information of a meta-graph.
Extensive experiments on two real datasets demonstrate that 
{\sun} and {\sunl} are more effective than state-of-the-art approaches.

\end{abstract}


\section{Introduction}

Many information retrieval and mining tasks such as node classification~\cite{chen2017task}, clustering~\cite{sun2013pathselclus}, link prediction~\cite{sun2012will}, and information diffusion \cite{sun2018multi} become time-consuming in large-scale networks. This motivates researchers to develop network embedding techniques which aim to learn a distributed representation vector for each node in a network.
An effective network embedding should preserve the similarity between nodes in order to reconstruct the original network.

The word2vec~\cite{mikolov2013efficient} idea has inspired many studies for network representation learning, most of which are in the context of homogeneous information networks, 
such as DeepWalk~\cite{perozzi2014deepwalk}, 
LINE~\cite{tang2015line}, and node2vec~\cite{grover2016node2vec}.
A homogeneous information network is a simple structural network,
where all nodes and links are considered to belong to a single class.

\begin{figure}[t]
    \centering
    \includegraphics[width=0.4\textwidth]{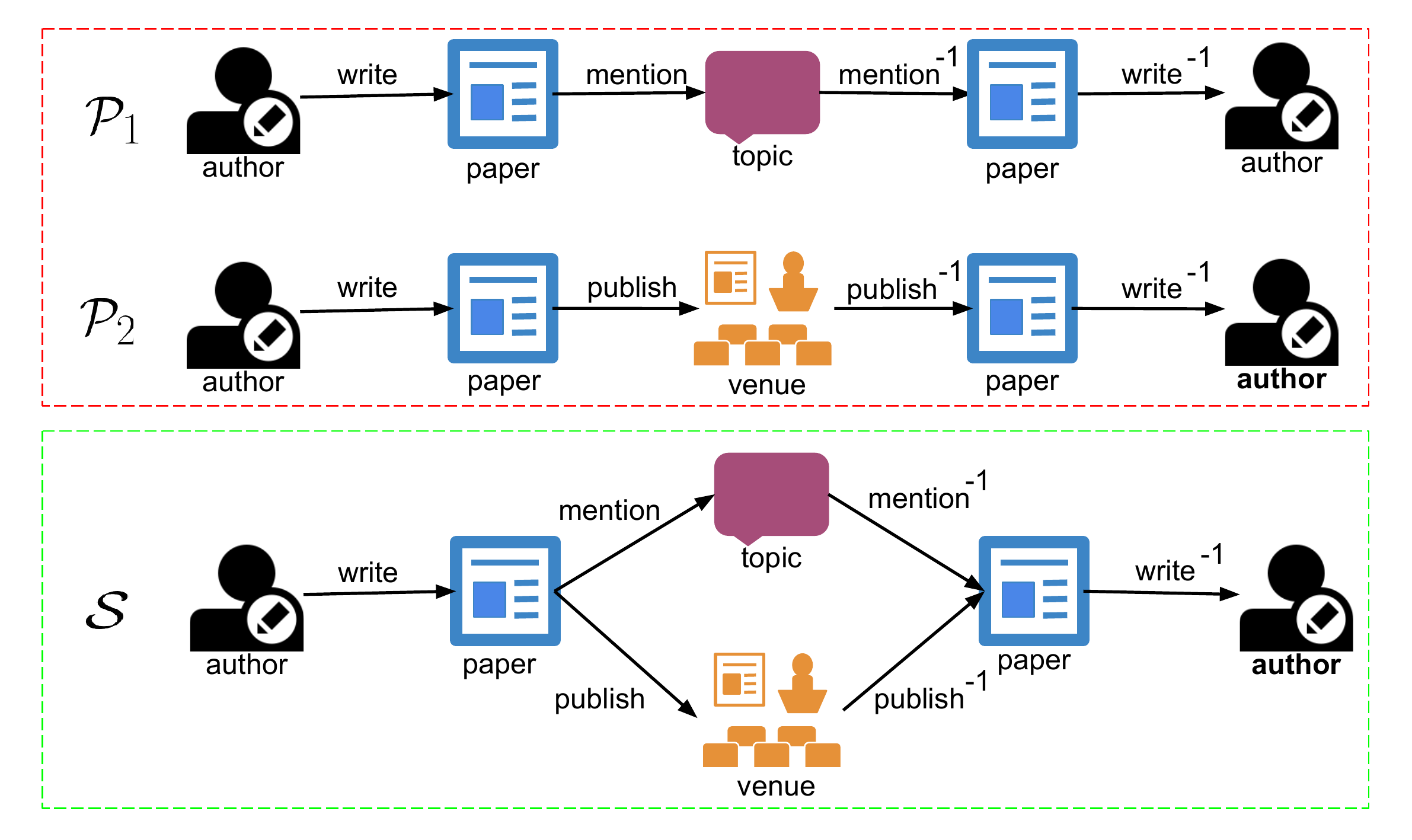}
    \caption{An example of meta-paths $\mathcal{P}_1$, $\mathcal{P}_2$, and meta-graph $\mathcal{S}$ of the bibliographic data.}
    \label{fig:schema}
\end{figure}

However, in practice, there are usually multiple types of nodes ({\em e.g.}, authors and papers in DBLP) and links ({\em e.g.}, cite and publish) that compose a heterogeneous information network (HIN).
To measure the similarity between nodes in HINs, many customized similarity or relevance measures based on meta-paths have been proposed in recent years ~\cite{lao2010relational,sun2011pathsim}.
For example, a meta-path $author \rightarrow paper \rightarrow venue \rightarrow paper \rightarrow author$ (denoted as $APVPA$) indicates two authors having their publications in the same venue.
Comparing to meta-path-based relevance measures utilizing only simple structural information, 
meta-graph~\cite{huang2016meta} is recently proposed to capture complex structural information in HINs. In short, meta-graph is a special directed acyclic graph (DAG) which contains at least two embedded meta-paths, such as a DAG containing $APVPA$ and $APTPA$ as shown in Figure \ref{fig:schema}, where $T$ is the topic of a paper. 

Meta-graph is an effective tool to calculate the relevance score between nodes in HINs, where a higher score indicates that there are more meta-graph instances between two nodes, {\em i.e.}, a closer relationship.
How to explore meta-graphs for representation learning in HINs is still an open question.
An intuitive idea for meta-graph-based representation learning is to learn the node embedding by leveraging multiple meta-graphs between nodes in HINs. However, existing meta-graph-based relevance measures only utilize the strong relations as defined by the meta-graphs themselves, and they usually ignore the weak relations as indicated by their embedded meta-paths.
To address this problem, we propose to learn the node embedding by leveraging both meta-graph and its embedded meta-paths for similarity search.
An effective representation learning based on a single meta-graph should contain both strong and weak relations embedded in this meta-graph.
In addition, we explore a novel meta-graph-based similarity measure to compute relevance scores that can better capture the strong relations between nodes in HINs.

In summary, there are three-fold contributions of this paper:
1) We are the first to propose the meta-graph-based node embedding method in HINs. Specifically, we develop two kinds of node embedding methods based on meta-graph, named {\sun} and {\sunl} respectively.
2) We introduce GraphSim which is an effective meta-graph-based similarity measure with best performance comparing to previous meta-graph-based similarity measures, such as StructCount and SCSE.
3) Our approaches show the best performance comparing to other competing methods on two real-world datasets.








\section{Preliminary and Problem Formulation}\label{sec:prelim}
In this section, we first introduce some related concepts and notations
from multilinear algebra.
Then, we review some concepts and approaches involved in HIN analysis
including meta-graph and relevance measure. 
Last part, we formulate the problem of node embedding in HINs.

\subsection{Multilinear Algebra}\label{sec:tensor}
The basic mathematical object of multilinear algebra is the tensor, a higher order generalization of vectors (first order tensors) and matrices (second order tensors) to multiple indices. The order of a tensor is the number of dimensions, also known as modes or ways. An $N$-th order tensor is represented as $\mathcal{X} \in \mathbb{R}^{I_{1} \times I_{2} \times \cdots \times I_{N}}$, where $I_n$ is the cardinality of its $n$-th mode, $n \in \{1,2, \cdots, N\}$. An element of a vector $\mathbf{x}$, a matrix $\mathbf{X}$, or a tensor $\mathcal{X}$ is denoted by $x_i$, $x_{i,j}$, $x_{i,j,k}$, etc., depending on the number of modes. All vectors are column vectors unless otherwise specified. For an arbitrary matrix $\mathbf{X} \in \mathbb{R}^{I \times J}$, its $i$-th row and $j$-th column vector are denoted by $\mathbf{x}^{i}$ and $\mathbf{x}_{j}$, respectively.


Definitions of outer product, partial symmetric tensor, mode-$n$ matricization, and CP factorization are given below, which will be applied to present our approach.
\begin{definition}(Outer Product) The outer product of $N$ vectors $\mathbf{x}^{(n)} \in \mathbb{R}^{I_n}$ for $n \in [1:N]$ is an $N$-th order tensor and defined element-wise by $\big (\mathbf{x}^{(1)} \circ \cdots \circ \mathbf{x}^{(N)}\big)_{i_1, \cdots, i_N} = x^{(1)}_{i_1}  \cdots x^{(N)}_{i_N}$ for all values of the indices.
\end{definition}


\begin{definition}(Partial Symmetric Tensor) An $N$-th order tensor is a rank-one partial symmetric tensor if it is partial symmetric on modes $i_1,...,i_j \in {1,...,N}$, and can be written as the tensor product of $N$ vectors, i.e.,
\begin{align}
\mathcal{X}={x}^{(1)} \circ \cdots \circ \mathbf{x}^{(N)}
\end{align}
where $x^{(i_1)}=\cdots=x^{(i_j)}$.
\end{definition}

\begin{definition}(Mode-$n$ Matricization) The mode-$n$ matricization or unfolding of an $N$-th order tensor $\mathcal{X} \in \mathbb{R}^{I_{1} \times I_{2} \times \cdots \times I_{N}}$ is denoted by $\mathbf{X}_{(n)}$ and is of size $I_n \times J_n $, where $J_n = \Pi_{m=1, m \neq n}^N I_m$.
\end{definition}


\begin{definition}(CP Factorization) For a general tensor $\mathcal{X}\in \mathbb{R}^{I_1\times\cdots\times I_N}$, its CANDECOMP/PARAFAC (CP) factorization is 
\begin{align}
\mathcal{X}= \sum_{r=1}^R \mathbf{x}_{r}^{(1)} \circ \cdots \circ \mathbf{x}_{r}^{(N)} = \llbracket \mathbf{X}^{(1)},...,\mathbf{X}^{(N)} \rrbracket
\end{align}
where for $n\in [1:N]$, $\mathbf{X}^{(n)}=[\mathbf{x}_1^{(n)},...,\mathbf{x}_R^{(n)}]$ are factor matrices of size $I_n \times R$, $R$ is the number of factors, and $\llbracket \cdot \rrbracket$ is used for shorthand.
\end{definition}

To obtain the CP factorization $\llbracket \mathbf{X}^{(1)}, \cdots, \mathbf{X}^{(N)} \rrbracket$, the objective is to minimize the following estimation error:
\begin{align}
     \mathcal{L} = \underset{\mathbf{X}^{(1)}, \cdots, \mathbf{X}^{(N)}}{\min} \| \mathcal{X} - \llbracket \mathbf{X}^{(1)}, \cdots, \mathbf{X}^{(N)} \rrbracket \|_F^2 \label{eq:CP_problem}
\end{align}
However, $\mathcal{L}$ is not jointly convex w.r.t. $\mathbf{X}^{(1)}, \cdots, \mathbf{X}^{(N)}$. A widely used optimization technique is the Alternating Least Squares (ALS) algorithm, which alternatively minimize $\mathcal{L}$ for each variable while fixing the other, that is,
\begin{equation}
\mathbf{X}^{(n)} \leftarrow \underset{\mathbf{X}^{(n)}}{\arg\min} \| \mathbf{X}_{(n)} - \mathbf{X}^{(n)} (\odot_{i \neq n}^N \mathbf{X}^{(i)} )^\mathrm{T} \|_F^2 \label{eq:ALS}
\end{equation}
where $\odot_{i \neq n}^N \mathbf{X}^{(i)} = \mathbf{X}^{(N)} \odot \cdots \mathbf{X}^{(n-1)} \odot \mathbf{X}^{(n+1)} \cdots \odot \mathbf{X}^{(1)}$.



\subsection{Meta Graph}\label{sec:HIN}

\begin{definition}\label{def:meta_structure}
(Meta-Graph~\cite{huang2016meta}) A meta-graph $S$ is a directed acyclic graph (DAG) defined on a HIN schema $T_G = (\mathcal{O}, \mathcal{R})$. A meta-graph $S$ contains a single source node $n_s$ with 0 in degree and a single target node $n_t$ with 0 out degree.
Mathematically, a meta-graph $S = (A, B, n_s, n_t)$, 
where $A$ is a set of nodes, $B$ is a set of edges, $n_s$ is the of source node,
and $n_t$ is the target node,. 
\end{definition}

Since a meta-graph only has one source node and one target node,
not all sub-graphs of HINs can be meta-graph.




\begin{definition}
(Meta-graph-based Relevance Measure) Given a HIN $G =(V, E)$ 
and a meta-graph $\mathcal{G}$,
the similarity of any two nodes $v_s, v_t \in V$ 
with respect to $\mathcal{G}$ is defined as:
\end{definition}

\begin{equation}\label{eq:pro}
    s = \sum_{g_{v_s \rightarrow v_t} \in  \mathcal{G}} s(v_s, v_t \; | \; g_{v_s \rightarrow v_t})
\end{equation}



where $g_{v_s \rightarrow v_t}$ is a meta-graph instance of $\mathcal{G}$, and 
$s(v_s, v_t \, | \, g_{v_s \rightarrow v_t})$ is the relevance score between $v_s$ and $v_t$, which will be determined by  the number of meta-graph instances connecting them.

Prior works provide different meta-graph-based relevance measures, such as 
{StructCount}, {SCSE} and {BSCSE}~\cite{huang2016meta}.

\subsection{Problem Formulation}\label{sec:embedding_def}

We study the problem of meta-graph-based node embedding in the HIN. 
Given a HIN $G=(V, E)$, we have two goals in this study. 
First, we want to explore a customized meta-graph-based relevance measure which can more efficiently capture the complex structural information. 
Second, we aim at finding an effective node embedding that can better preserve the closeness between nodes in a HIN based on a meta-graph and its embeded meta-paths analysis. 
Specifically, we integrate all the similarity information of a meta-graph and its embedded meta-paths into a symmetric matrix and a partial symmetric tensor, 
and perform multilinear analysis of the coupled partial symmetric tensor 
and symmetric matrix to find the node embedding.

\begin{figure*}
    \centering
    \includegraphics[width=0.9 \textwidth]{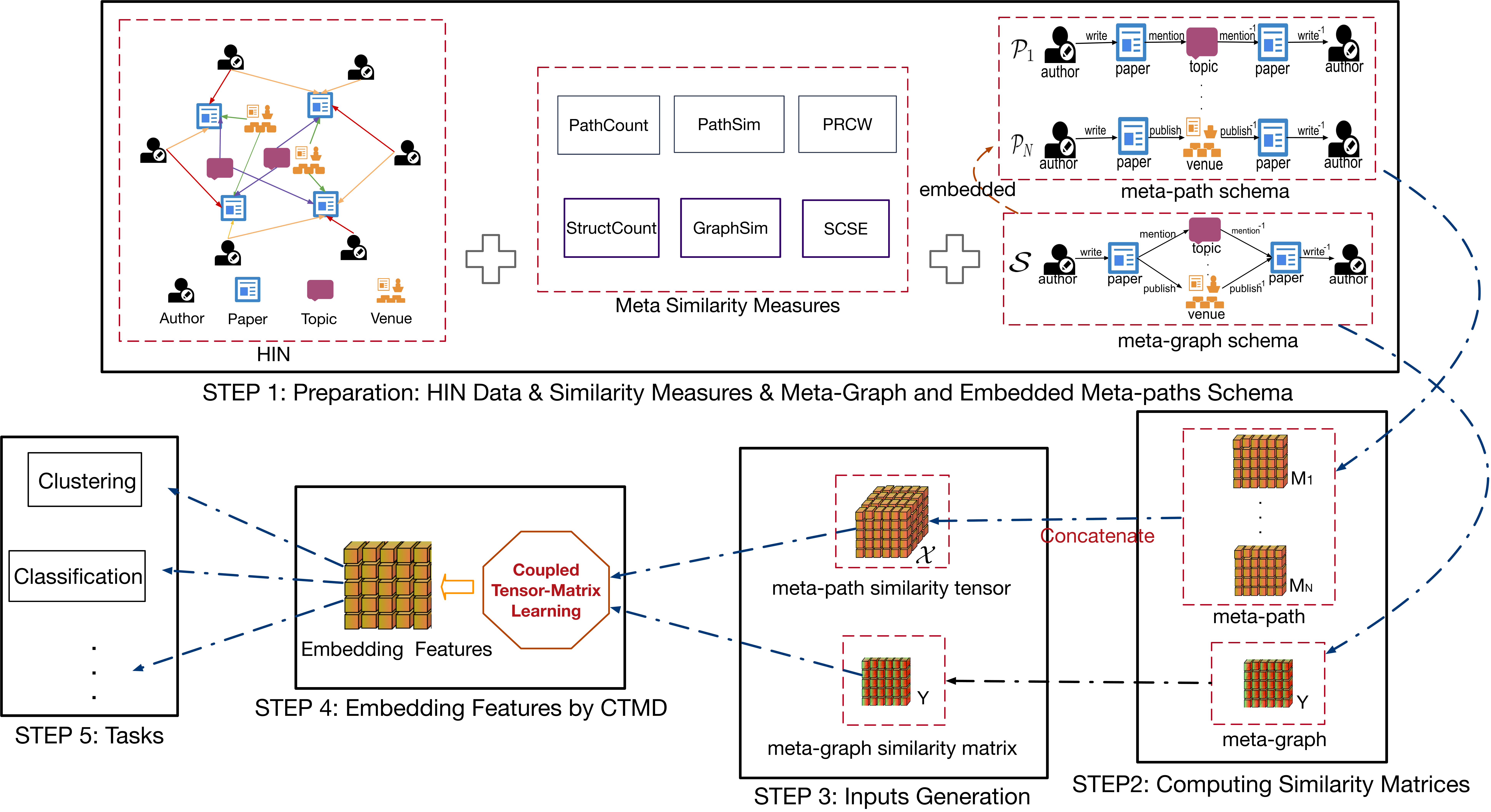}
    \caption{Framework of Meta-graph-based Node Embedding with Coupled Tensor-Matrix Decomposition}
    \label{fig:tensor}
\end{figure*}

\section{Methods}\label{sec:methods}
In this section, we will introduce a brand new similarity measure,
and the embedding techniques of {\sunl}.
First, we will introduce a meta-graph-based similarity measure named
GraphSim. Then, we proposed a coupled tensor-matrix decomposition
to obtain a joint embedding for nodes in HINs.

\subsection{GraphSim: A Normalized version of StructCount}
First, we want to propose a new meta-graph-based similarity measure called
GraphSim. In previous work, Huang et al.~\cite{huang2016meta} proposed three
meta-graph-based similarity measures: StructCount, SCSE, 
and BSCSE which is a mixed measure based on previous two measures. 
GraphSim can be viewed as a normalized version of StructCount.

StructCount~\cite{huang2016meta} is a straightforward 
meta-graph-based similarity measure in HIN, 
which counts the number of meta-graph instances 
in the graph $G$ with an $n_s$ as source and an $n_t$ as target object. 





\begin{definition}
(GraphSim) A meta-graph-based similarity measure.
Given a symmetric meta-graph $\mathcal{G}$, GraphSim between two nodes
$v_t, v_s \in V$ is defined as:
\begin{equation}\label{eq:GraphSim}
    s(v_t, v_s) = \frac{2 \times |\{g_{v_t \rightarrow v_s}
    |g_{v_t \rightarrow v_s \in \mathcal{G}}\}|}{|\{g_{v_t \rightarrow v_t}
    |g_{v_t \rightarrow v_t \in \mathcal{G}}\}| + |\{g_{v_s \rightarrow v_s}
    |g_{v_s \rightarrow v_s\in \mathcal{G}}\}|}
\end{equation}
where $g_{v_t \rightarrow v_s}$is a meta-graph instance 
between $v_t$ and $v_s$, $g_{v_t \rightarrow v_t}$ is that between
$v_t$ and $v_t$, and $g_{v_s \rightarrow v_s}$ is that between $v_s$ and $v_s$.
\end{definition}



Comparing to StructCount, GraphSim is normalized version of StructCount.
$s_{GraphSim}(v_t, v_s)$ is determined by two parts:
First, the number of meta-graph instance between $v_t, v_s \in V$ by
following $\mathcal{G}$; Second, the balance of their visibility,
where the visibility is defined as the number of meta-graph instances 
between themselves. Normalized relevance score can present better relation between different nodes. For example, an author $A_1$ published all his four papers with $A_2$.
$A_3$ published five papers with $A_2$, and $A_3$ published other five papers with other authors. Without normalized process, the relation between $A_2$ and $A_3$ is closer than $A_2$ and $A_1$. However, for common sense, we should agree $A_2$ and $A_1$ have closer relation, which indicates GraphSim is a better measure.

Comparing to three measures in~\cite{huang2016meta}, 
$s(v_t, v_s)$ of GraphSim is between 0 and 1 like SCSE. 
However, SCSE measures the random walk probability that $v_s$ expands a meta-graph instance to $v_t$. 
In our application, we find GraphSim shows better performance than all those three meta-graph measures~\cite{huang2016meta}.

\subsection{{\sunl}: Node Embedding by CTMD}
In this section, we show how to jointly consider similarity information of a meta-graph and its embedded meta-paths to learn a node embedding. The basic idea is the integration of similarity matrices and coupled embedding by joint factorization. Specifically, we first compute a meta-graph similarity matrix using the proposed GraphSim, denoted as $\mathbf{Y} \in \mathbb{R}^{M \times M}$, and for each meta-path $\mathcal{P}$, compute an embedded meta-path similarity matrix using the PathSim, denoted as $\mathbf{M} \in \mathbb{R}^{M \times M}$. Next, we concatenate the embedded meta-path similarity matrices of different embedded meta-paths to form a third-order tensor comprising three modes: nodes, nodes, and paths, denoted as $\mathcal{X} = [\mathbf{M}_1, \cdots, \mathbf{M}_N]$ $\in \mathbb{R}^{M \times M \times N}$. Then, we introduce a novel coupled tensor-matrix decomposition (CTMD) method to find common latent features between $\mathcal{X}$ and $\mathbf{Y}$. Last, we use the latent features to measure the similarity between different nodes in the HIN. 

In the following, we detail the CTMD method, which can be seen as a special case of the coupled tensor-matrix decomposition \cite{acar2011all} with input partial symmetric tensor $\mathcal{X}$ and symmetric matrix $\mathbf{Y}$. Notice that since similarity matrix is symmetric, the resulting $\mathcal{X}$ is a partial symmetric tensor, and $\mathbf{Y}$ is a symmetric matrix.

Tensors (including matrix) provide a natural and efficient representation for a meta-graph data, but there is no guarantee that such representation will be good for subsequent learning, since learning will only be successful if the regularities that underlie the data can be discerned by the model. Tensor factorization is a powerful tool to analyze tensors. In previous work, it was found that CP factorization (which is a higher order generalization of SVD) is particularly effective to acknowledge the connections and find valuable features among tensor data \cite{van2016structured}. Motivated by these observations, we exploit the benefits of CP and SVD factorizations to find an effective embedding in the sense of meta-path-based similarity tensor $\mathcal{X}$ and meta-graph similarity matrix $\mathbf{Y}$.

Based on above analysis, we design our CTMD objective function as below:
\vspace{-2pt}
\begin{align}
\underset{\mathbf{P}, \mathbf{T}}{\min}~~ \| \mathcal{X} - \llbracket \mathbf{P}, 
\mathbf{P}, \mathbf{T} \rrbracket \|_F^2 + \alpha \|\mathbf{Y} - \mathbf{P} 
\mathbf{P}^\mathrm{T} \|_F^2
\label{eq:obj}
\end{align}
\vspace{-2pt}
where $\mathbf{P} \in \mathbb{R}^{M \times R}$ and $\mathbf{T} \in \mathbb{R}^{N \times R}$ are latent matrices. Specifically, $\mathbf{P}$ is jointly learned from both meta-graph and meta-path similarity information.


The objective function in Eq.~(\ref{eq:obj}) is non-convex with respect to $\mathbf{P}$ and $\mathbf{T}$ together, thus there is no closed-form solution. We introduce an effective iteration method to solve this problem. The main idea is to decouple the parameters using an Alternating Direction Method of Multipliers (ADMM) approach \cite{boyd2011distributed}, by alternatively optimizing the objective with respect to one variable, while fixing others.

\textbf{Update $\mathbf{P}$}: First, we optimize $\mathbf{P}$ while fixing $\mathbf{T}$. Notice that the objective function in Eq.~(\ref{eq:obj}) involves a fourth-order term with respect to $\mathbf{P}$ which is difficult to optimize directly. To obviate this problem, we use a variable substitution technique and minimize the following objective function
\begin{align}
& \underset{\mathbf{P}, \mathbf{Q}}{\min}~~ \| \mathcal{X} - \llbracket \mathbf{P}, \mathbf{Q}, \mathbf{T} \rrbracket \|_F^2 + \alpha \|\mathbf{Y} - \mathbf{P} \mathbf{Q}^\mathrm{T} \|_F^2 \nonumber \\
& s.t.~~ \mathbf{P} = \mathbf{Q}
\label{eq:obj_trans}
\end{align}
where $\mathbf{Q} \in \mathbb{R}^{M \times R}$ is an auxiliary variable.

The augmented Lagrangian function of Eq.~(\ref{eq:obj_trans}) is
\begin{align}
\mathcal{L}(\mathbf{P}, \mathbf{Q}) = & \| \mathcal{X} - \llbracket \mathbf{P}, \mathbf{Q}, \mathbf{T} \rrbracket \|_F^2 + \alpha \|\mathbf{Y} - \mathbf{P} \mathbf{Q}^\mathrm{T} \|_F^2  \nonumber \\
& + tr(\mathbf{U}^\mathrm{T} (\mathbf{P} - \mathbf{Q})) + \frac{\lambda}{2}\| \mathbf{P} - \mathbf{Q} \|_F^2
\label{eq:obj_trans_L}
\end{align}
where $\mathbf{U} \in \mathbb{R}^{M \times R}$ is the Lagrange multiplier, and $\lambda$ is the penalty parameter which can be adjusted efficiently according to \cite{lin2011linearized}.

To compute $\mathbf{P}$, Eq.~(\ref{eq:obj_trans_L}) can be transformed as
\begin{align}
\underset{\mathbf{P}}{\min} \| \mathcal{X}_{(1)} - \mathbf{P} \mathbf{F}^\mathrm{T} \|_F^2 + \alpha \|\mathbf{Y} - \mathbf{P} \mathbf{Q}^\mathrm{T} \|_F^2 + \frac{\lambda}{2} \| \mathbf{P} - \mathbf{Q} + \frac{1}{\lambda} \mathbf{U} \|_F^2
\label{eq:optimize_P}
\end{align}
where $\mathbf{X}_{(1)} \in \mathbb{R}^{M \times (MN)}$ 
is the mode-1 matricization of $\mathcal{X}$, 
and $\mathbf{F} = \mathbf{T} \odot \mathbf{Q} \in \mathbb{R}^{(MN) \times R}$.

By setting the derivative of Eq.~(\ref{eq:optimize_P}) with respect to $\mathbf{P}$ to zero, we obtain the closed-form solution
\begin{align}
\mathbf{P} = (2\mathbf{X}_{(1)} \mathbf{F} + 2 \alpha \mathbf{Y} \mathbf{Q} + \lambda \mathbf{Q} - \mathbf{U})(2 \mathbf{F}^\mathrm{T} \mathbf{F} + 2 \alpha \mathbf{Q}^\mathrm{T} \mathbf{Q} + \lambda \mathbf{I})^{-1}
\label{eq:solution_P}
\end{align}
To efficiently compute $\mathbf{F}^\mathrm{T} \mathbf{F}$, we consider the following property of the Khatri-Rao product of two matrices
\begin{align}
\mathbf{F}^\mathrm{T} \mathbf{F} = (\mathbf{T} \odot \mathbf{Q})^\mathrm{T} (\mathbf{T} \odot \mathbf{Q}) = \mathbf{T}^\mathrm{T} \mathbf{T} \ast  \mathbf{Q}^\mathrm{T} \mathbf{Q}
\label{eq:property_kr}
\end{align}
Then the auxiliary matrix $\mathbf{Q}$ can be optimized successively in a similar way, and the solution is
\begin{align}
\mathbf{Q} = (2\mathbf{X}_{(2)} \mathbf{G} + 2 \alpha \mathbf{Y}^\mathrm{T} \mathbf{P} + \lambda \mathbf{P} + \mathbf{U})(2 \mathbf{G}^\mathrm{T} \mathbf{G} + 2 \alpha \mathbf{P}^\mathrm{T} \mathbf{P} + \lambda \mathbf{I})^{-1}
\label{eq:solution_Q}
\end{align}
where $\mathbf{X}_{(2)}$ is the mode-2 matricization of $\mathcal{X}$, and $\mathbf{G} =\mathbf{T} \odot \mathbf{P}$.

Moreover, we optimize the Lagrange multiplier $\mathbf{U}$ using the gradient descent method by
\begin{align}
\mathbf{U} \leftarrow \mathbf{U} + \lambda (\mathbf{P} - \mathbf{Q})
\label{eq:update_U}
\end{align}

\textbf{Update $\mathbf{T}$}: Next, we optimize $\mathbf{T}$ while fixing $\mathbf{P}$ and $\mathbf{S}$. We need to optimize the following objective function
\begin{align}
\underset{\mathbf{T}}{\min}~~ \| \mathbf{X}_{(3)} - \mathbf{T} \mathbf{H}^\mathrm{T}\|_F^2
\label{eq:optimize_T}
\end{align}
where $\mathbf{X}_{(3)}$ is the mode-3 matricization of $\mathcal{X}$, and $\mathbf{H} =\mathbf{Q} \odot \mathbf{P}$.

By setting the derivative of Eq.~(\ref{eq:optimize_T}) with respect to $\mathbf{T}$ to zero, we obtain the closed-form solution as
\begin{align}
\mathbf{T} = (\mathbf{X}_{(3)} \mathbf{H})(\mathbf{H}^\mathrm{T} \mathbf{H})^{-1}
\label{eq:solution_T}
\end{align}

The overall algorithm is summarized in Algorithm \ref{algo:CTMD}.

\begin{algorithm}
  \caption{Coupled Tensor-Matrix Decomposition (CTMD)}\label{algo:CTMD}
  \begin{algorithmic}[1]
    \REQUIRE Meta-path similarity tensor $\mathcal{X}$, and meta-graph similarity matrix $\mathbf{Y}$
    \ENSURE Embedding matrix $\mathbf{P}$
    \STATE: Set $\lambda_{max} = 10^{6}$, $\rho = 1.15$ 
    \STATE: Initialize $ \mathbf{P}, \mathbf{Q}, \mathbf{T}  \sim\mathcal{N}(0,1), \mathbf{U}=\mathbf{0}, \lambda=10^{-6}$    
    \ALOOP {convergence} 
        \STATE: Update P by Eq.~(\ref{eq:solution_P})
        \STATE: Update Q by Eq.~(\ref{eq:solution_Q})
        \STATE: Update T by Eq.~(\ref{eq:solution_T})
        \STATE: Update $\mu$ by $\mu$ $\leftarrow$ min($\rho\mu$, $\mu_{max}$)
    \ENDALOOP
  \end{algorithmic}
\end{algorithm}

\subsection{Time Complexity}
Each iteration in Algorithm \ref{algo:CTMD} consists of simple matrix operations. 
Therefore, rough estimates of its computational complexity 
can be easily derived based on ADMM ~\cite{liavas2015parallel}.

The estimate for the update of $\mathbf{P}$ according to 
 Eq.~(\ref{eq:solution_P}) is as follows: $O(M^2NR)$ for the computation
of the term $2\mathbf{X}_{(1)} \mathbf{F} + 2 \alpha \mathbf{Y} \mathbf{Q} + \lambda \mathbf{Q} - \mathbf{U}$; 
$O((M + N) R^2)$ for the computation of the term $2 \mathbf{F}^\mathrm{T} \mathbf{F} + 2 \alpha \mathbf{Q}^\mathrm{T} \mathbf{Q} + \lambda \mathbf{I}$ due to Eq.~(\ref{eq:property_kr}) and $O(R^3)$ for its Cholesky decomposition;
$O(MK^2)$ for the computation of the system solution that gives the
updated value of $\mathbf{P}$. An analogous estimate can be
derived for the update of $\mathbf{Q}$.

Overall, the updates of model parameters $\mathbf{P}$ and $\mathbf{Q}$, 
require O($R^3 + (M+N)R^2 + M^2NR$) arithmetic
operations in total. 

\begin{algorithm}
  \caption{{\sunl}}\label{algo:metagraph2vec}
  \begin{algorithmic}[1]
    \REQUIRE An HIN G, a particular meta-graph $g$, a embedded meta-path $p_i$ of a meta-graph $g$, and an empty array $\mathbf{P}_A$ 
    \ENSURE Embedding matrix $\mathbf{P}$
    \STATE: Y = GraphSim(G, $g$)
    \ALOOP {$p_i \in g$} 
        \STATE: $\mathbf{P}_i$ = PathSim(G, $p_i$)
        \STATE: $\mathbf{P}_A$ = [$\mathbf{P}_A$; $\mathbf{P}_i$] store $\mathbf{P}_i$ into $\mathbf{P}_A$
    \ENDALOOP
    \STATE: $\mathcal{X}$ = concatenate($\mathbf{P}_A$) $\mathcal{X}$ is a tensor of $\mathcal{S}$'s embedded meta-paths
    \STATE: P = CTMD($\mathcal{X}$, Y)
    
  \end{algorithmic}
\end{algorithm}



\section{Experiments}\label{sec:experiments}


In this section, we conduct extensive experiments in order to
test the effectiveness of the proposed methods:
GraphSim, {\sun} and {\sunl}.
We first introduce two real-life datasets and a set of methods to be compared. 
Then, we evaluate the effectiveness of proposed methods
on four data mining tasks: 
clustering, classification, 
parameter analysis and time analysis. 

\begin{figure}[]
    \centering
    \includegraphics[width=0.4\textwidth]{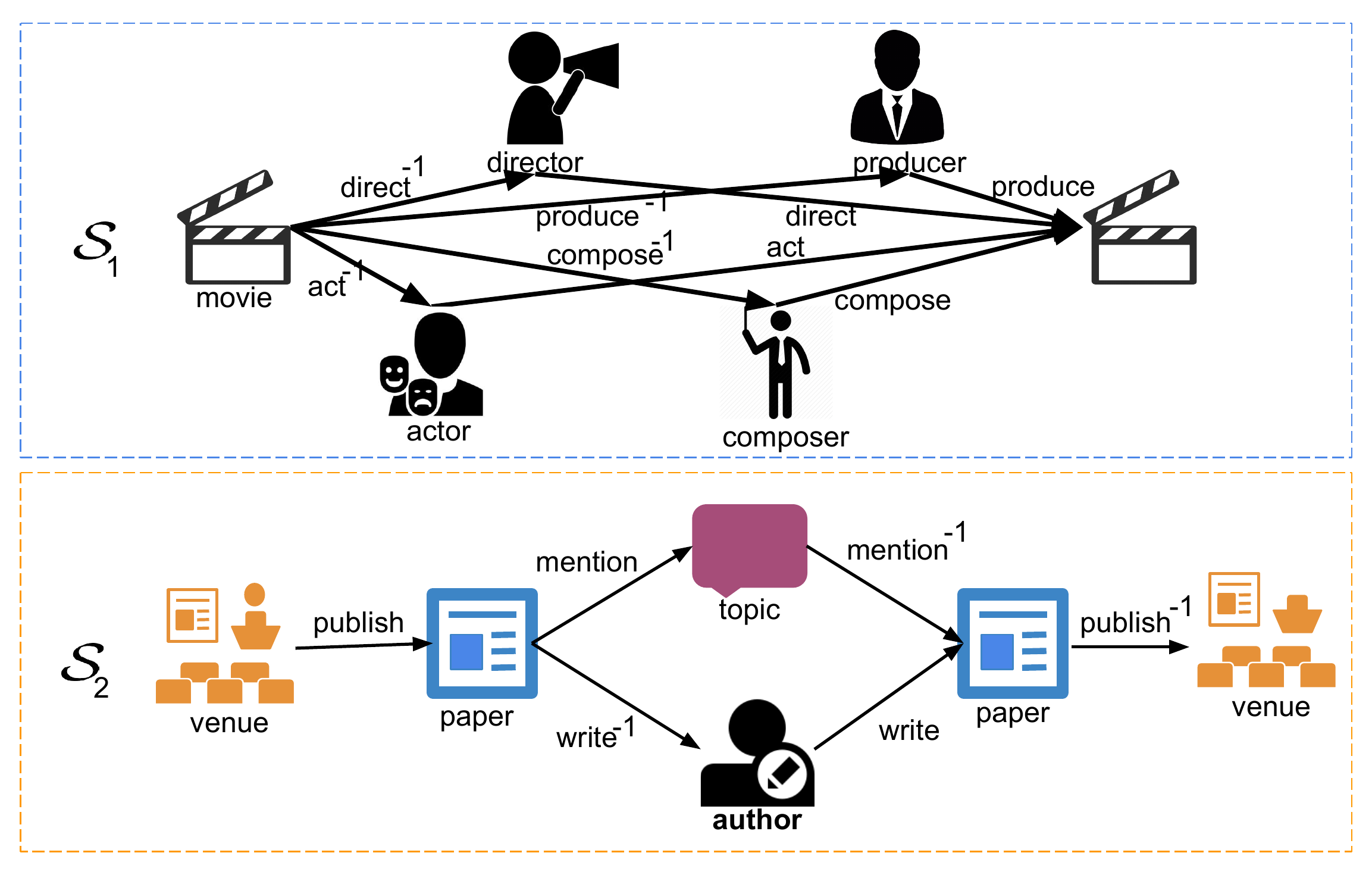}
    \caption{Guided meta-graphs: $\mathcal{S}_1$ is the guided meta-graph for M2M task, and $\mathcal{S}_2$ is for V2V task.  }
    \label{fig:meta-graph}
\end{figure}

We use two real datasets ({\em e.g.} DBLP-4-Area and YAGO Movie) in the evaluation. Table~\ref{table:statistics} shows some statistics about them.
DBLP-4-Area \cite{sun2011pathsim} is the subset of original DBLP, 
which contains 5,237 papers (P), 5,915 authors (A), 18 venues (V), 4,479 topics (T). 
The authors and venues are from 4 areas: \emph{database}, \emph{data mining}, \emph{machine learning} and \emph{information retrieval}.
YAGO Movie is a subset of YAGO \cite{huang2016meta}, which contains 7,332 movies (M), 10,789 actors (A), 1,741 directors (D), 3,392 producers (P) and 1,483 composers (C). 
The movies are divided into five genres: \emph{action}, \emph{horror}, \emph{adventure}, \emph{sci-fi} and \emph{crime}. The guided meta-graphs are designed for three tasks as shown in the Figures \ref{fig:schema} and \ref{fig:meta-graph}.

The proposed methods are compared with meta-graph-based relevance measures ({\em e.g.} StructCount, SCSE, and BSCSE \cite{huang2016meta}), 
and network embedding approaches ({\em e.g.} DeepWalk \cite{perozzi2014deepwalk}, and LINE \cite{tang2015line}) in clustering and classification tasks. The experimental results are shown in the following sections.


\begin{table}[] \small
\centering
\caption{Statistics of Datasets}
\label{table:statistics}
\begin{tabular}{|l|c|c|c|c|c|}
\hline
      & $|V|$  & $|E|$  & Avg. degree & $|\mathcal{L}|$ & $|\mathcal{R}|$ \\  \hline
DBLP  & 15,649 & 51,377 & 6.57        & 4              & 4            \\ \hline 
MOVIE & 25,643 & 40,173 & 3.13        & 5              & 4            \\ \hline 

\end{tabular}
\end{table}

\begin{table*}[] \small
\centering
\caption{Clustering performance}
\label{table:clustering}
\resizebox{7in}{!}{%
\begin{tabular}{|l|l|c|c|c|c|c|c|c|c|}
\hline
  &   & \multicolumn{3}{c|}{Pre. Meta-Graph Measures} & \multicolumn{2}{c|}{Pre. Network Embedding} & \multicolumn{3}{c|}{OUR WROKS} \\ \hline 
Task  & Method   & StuctCount    & SCSE       & BSCSE($\alpha=1$)      & LINE$^*$   & DeepWalk$^*$             & GraphSim  & {\sun}      & {\sunl}     \\ \hline 
\multirow{2}{*}{\begin{tabular}[c]{@{}l@{}}DBLP\\ (V2V)\end{tabular}} 
  & NMI      & 0.2634        & 0.6309     & 0.6309     & 0.7954             & 0.8258               & 0.8479     & 0.8521  & \textbf{0.8718} \\ 
  & Purity   & 0.5000        & 0.7333     & 0.7333     & 0.8042             & 0.8584               & 0.8744     & 0.8817  & \textbf{0.8956}  \\ \hline 
\multirow{2}{*}{\begin{tabular}[c]{@{}l@{}}DBLP\\ (A2A)\end{tabular}} 
  & NMI      & 0.0338   & 0.0156    & 0.0156        & 0.3920    & 0.4896        & 0.2150     & 0.5263  & \textbf{0.5315}  \\ 
  & Purity   & 0.2997   & 0.2822    & 0.2823        & 0.7135    & 0.7941        & 0.4903     & 0.7956  & \textbf{0.7989}  \\ \hline 
\multirow{2}{*}{\begin{tabular}[c]{@{}l@{}}Moive\\ (M2M)\end{tabular}} 
  & NMI      & 0.0011   & 0.0008    & 0.0008        & 0.0008    & 0.0007        & 0.0021     & 0.0045  & \textbf{0.0045}  \\ 
  & Purity   & 0.2991   & 0.2988    & 0.2988        & 0.2981    & 0.2981        & 0.3002     & 0.3017  & \textbf{0.3032}  \\ \hline 
\multirow{2}{*}{\begin{tabular}[c]{@{}l@{}}Overall\end{tabular}}
  & NMI    & 0.0994   & 0.2158    & 0.2158        & 0.3961    & 0.4387        & 0.3550     & 0.4610  & \textbf{0.4693}  \\ 
  & Purity & 0.3663   & 0.4381    & 0.4381        & 0.6052    & 0.6502        & 0.5550     & 0.6597  & \textbf{0.6659}  \\ \hline
\end{tabular}
}
\end{table*}

\begin{table*}[] \small
\centering
\caption{Classification performance}
\label{talbe:classification}
\resizebox{7in}{!}{%
\begin{tabular}{|l|l|c|c|c|c|c|c|c|c|}
\hline
&  & \multicolumn{3}{c|}{Pre. Meta-Graph Measures} & \multicolumn{2}{c|}{Pre. Network Embedding} & \multicolumn{3}{c|}{OUR WROKS} \\ \hline 
Task    & Method   & StuctCount    & SCSE       & BSCSE($\alpha=0$)      & LINE$^*$ & DeepWalk$^*$  & GraphSim  & {\sun}      & {\sunl}     \\ \hline 
\multirow{2}{*}{\begin{tabular}[c]{@{}l@{}}DBLP\\ (A2A)\end{tabular}} 
& Macro-F1      & 0.734      & 0.616      & 0.734       & 0.816   & 0.839     & 0.818     & 0.863  & \textbf{0.867} \\   
& Micro-F1      & 0.730      & 0.634      & 0.730       & 0.817   & 0.840     & 0.819    & 0.863 & \textbf{0.867} \\ \hline 
\multirow{2}{*}{\begin{tabular}[c]{@{}l@{}}Movie\\ (M2M)\end{tabular}} 
& Macro-F1      & 0.126    & 0.111    & 0.126    & 0.186   & 0.189     &  0.125    & 0.298  & \textbf{0.310} \\   
& Micro-F1      & 0.281    & 0.276    & 0.281    & 0.241   & 0.245     & 0.307     & 0.342 & \textbf{0.352} \\ \hline 
\multirow{2}{*}{\begin{tabular}[c]{@{}l@{}}Overall \end{tabular}} 
& Macro-F1      & 0.430   & 0.364    & 0.430        & 0.501    & 0.514        & 0.472     & 0.581  & \textbf{0.589}  \\ 
& Micro-F1      & 0.506   & 0.455    & 0.506        & 0.540    & 0.543        & 0.563     & 0.603  & \textbf{0.610}\\ \hline
\end{tabular}
}
\end{table*}

\subsection{Clustering Results}\label{sec:clustering}


We first conduct a clustering task to evaluate the performance of the compared methods on DBLP and YAGO Movie datasets.
For DBLP, we use the areas of authors as ground-truth label for clustering authors (A2A),
and use the areas of venues as labels for clustering venues (V2V). 
For YAGO Movie, we use the genres of movies as labels (M2M). 
To be specific, we use $k$-means on the derived meta-graph-based relevance matrices for the clustering task.
To evaluate the results, we use NMI and purity as evaluation metrics.

Clustering results of the three tasks are shown in Table~\ref{table:clustering}.
Comparing to previous meta-graph-based relevance measures,
the proposed GraphSim always shows the best performance of all.
We observe at least 19.94\% improvement in NMI of GraphSim method when compared with the previous meta-graph-based relevance measure on clustering the venues and authors in DBLP, respectively.
The clustering results can be sensitive to initialization of centroid seeds, so we set 100 times of random initializations.
All methods show worse performance on YAGO Movie than DBLP,
but the proposed methods, especially {\sunl}, show the best performance comparing to prior works..
\vspace{-2pt}
\subsection{Classification Results}\label{sec:classification}
We then conduct a classification task. 
Comparing to the clustering task, in DBLP we do not evaluate the results of classifying
the venues, as the total number of venues is only 18.
We first apply previous methods and our works to generate the similarity matrices or
embedding space of the original network. Then, we randomly partition the samples,
and set 80\% samples as training set and the rest as testing set.
Last, we apply $k$ nearest neighbor (k-NN) classifier with $k = 5$ to evaluate the methods
with training and testing dataset~\cite{sun2011pathsim,huang2016meta}. 
To prevent the special case of random partition, 
we repeat and use different random partition 10 times in total.
For multi-label classification task, we use the average Macro-F1 score
and Micro-F1 score as the evaluation metrics.


GraphSim outperforms the existing relevance measures ({\em e.g} StructCount, SCSE and BSCSE) because it represents a better relations between objects in the HINs by normalizing the presence of meta-graph structures. {\sunl} outperforms all the baselines because it captures both lower-order ({\em i.e.} meta-path) and higher-order ({\em i.e.} meta-graph) structural information by facilitating the use of coupled tensor-matrix decomposition method to obtain a joint embedding for nodes in HINs.

\begin{figure}[t] \small
\centering
\begin{tabular}{cc}
\subfloat[]{\includegraphics[width=.19\textwidth]{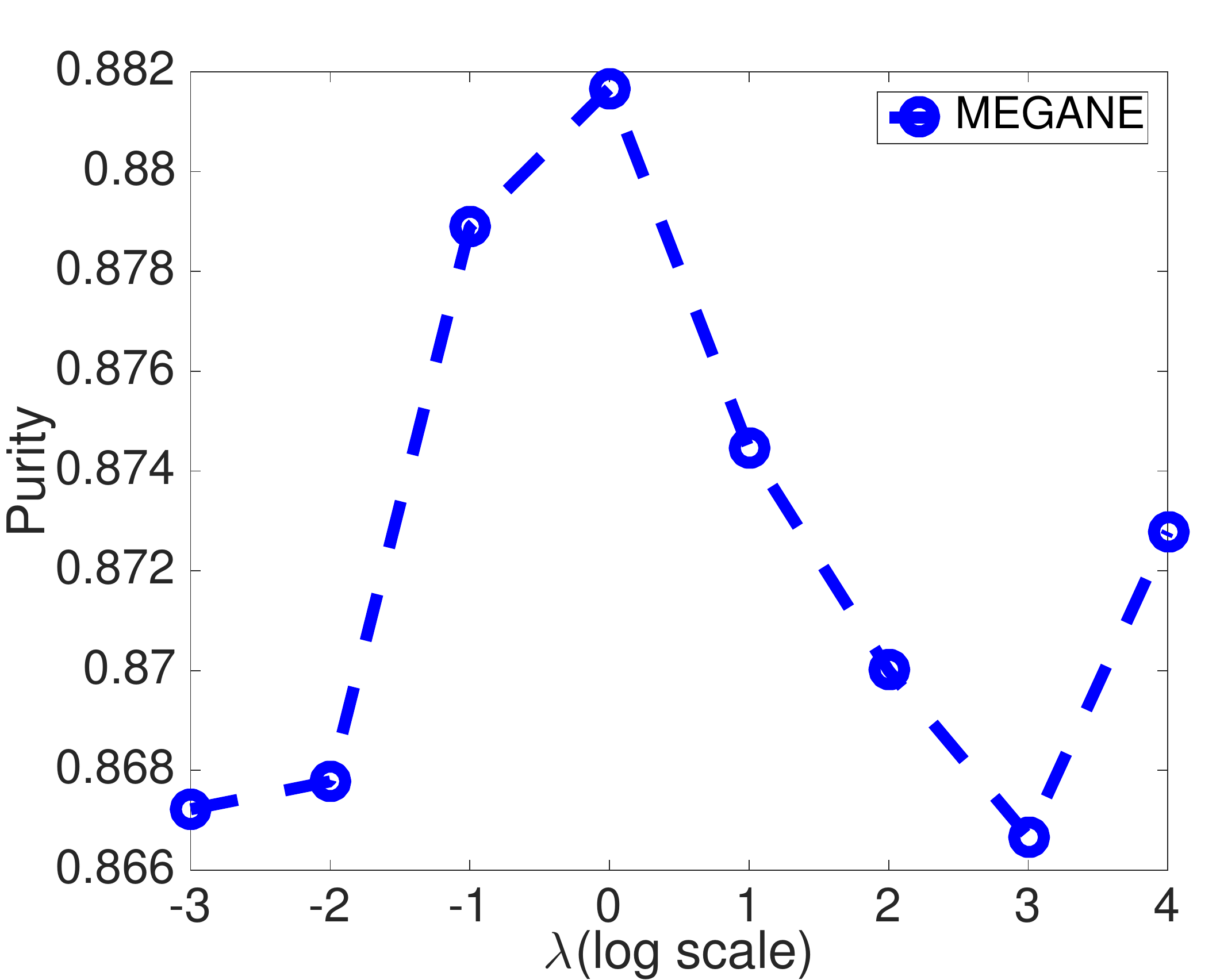}}
& \subfloat[]{\includegraphics[width=.19\textwidth]{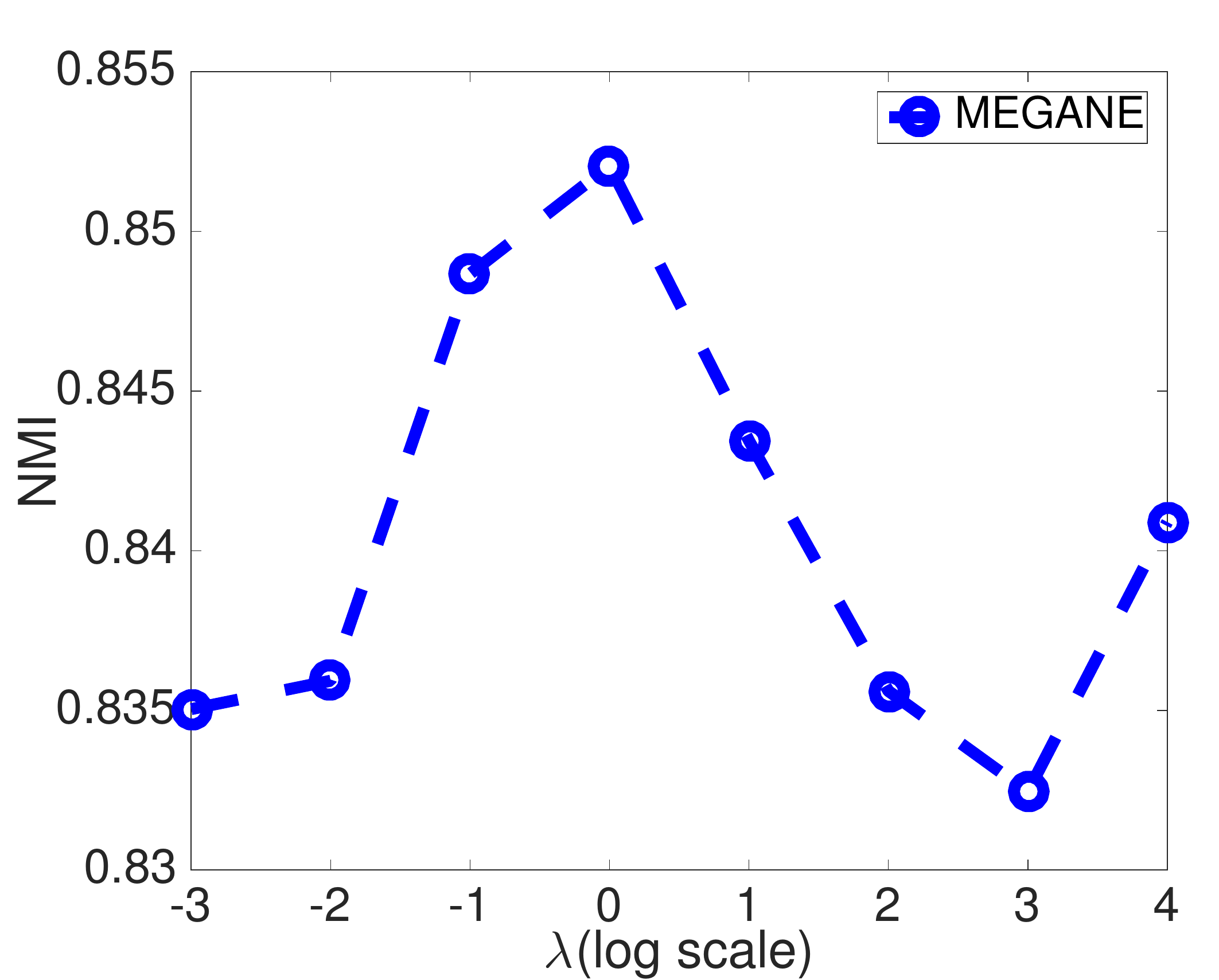}} \\
\subfloat[]{\includegraphics[width=.19\textwidth]{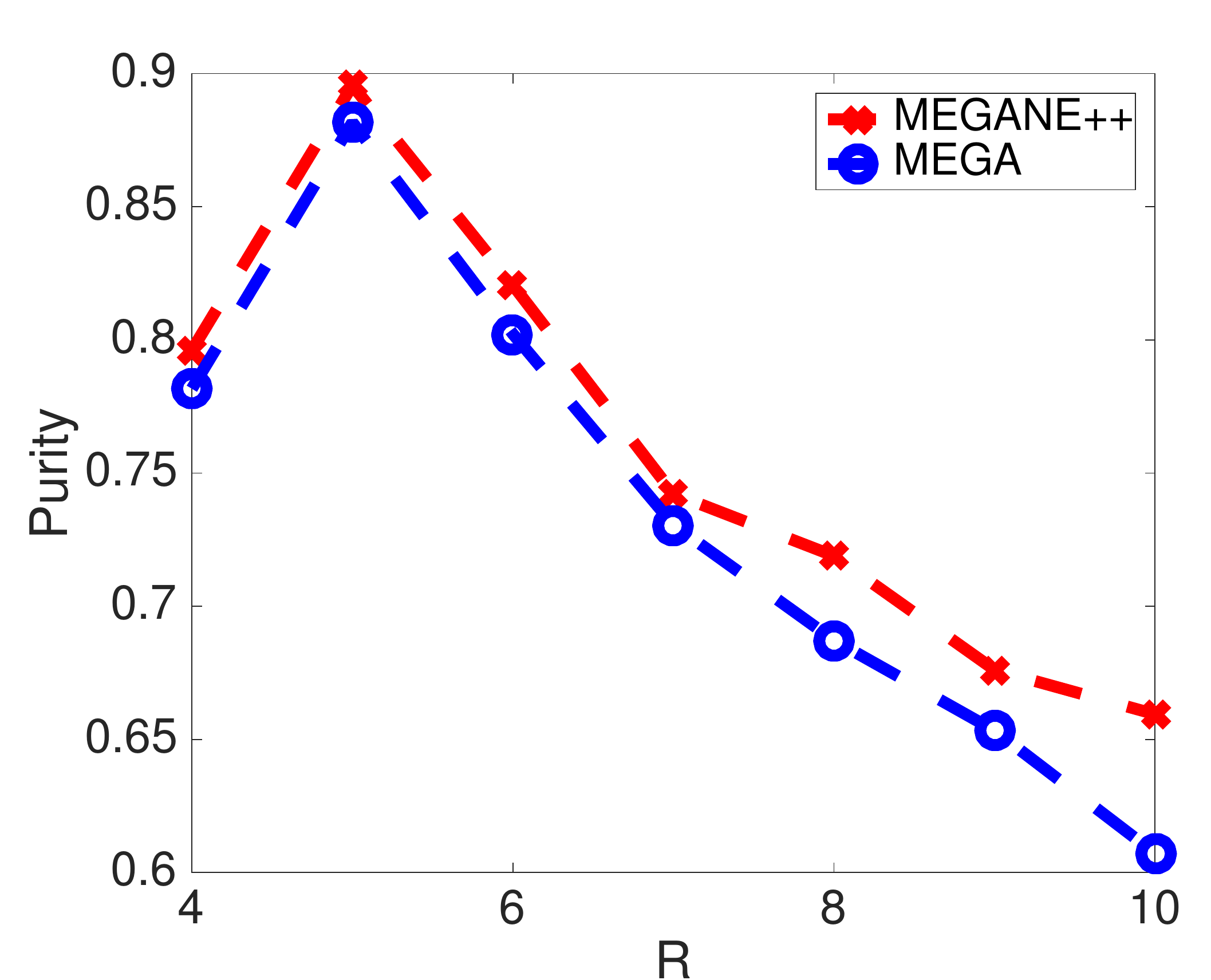}} 
& \subfloat[]{\includegraphics[width=.19\textwidth]{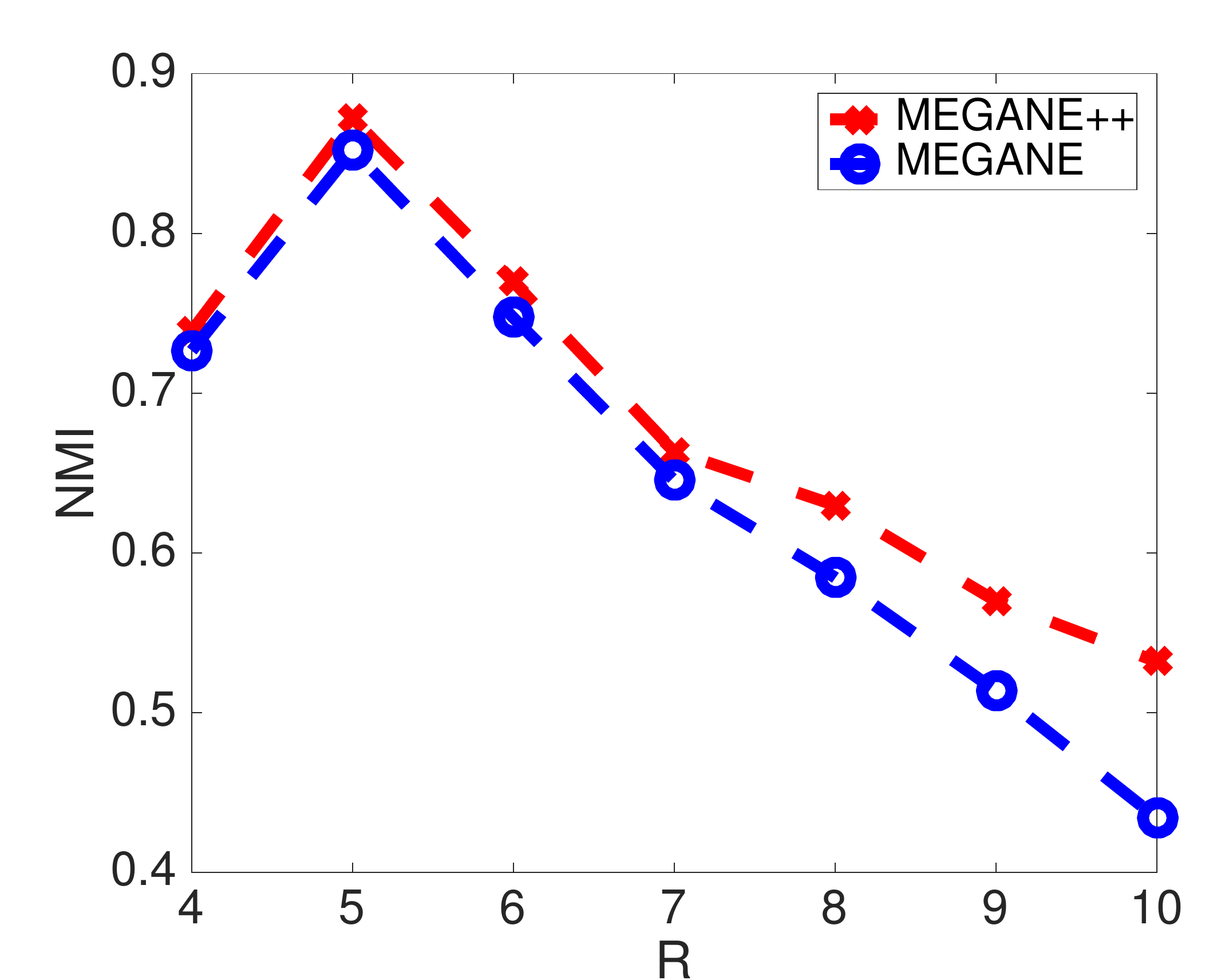}} \\
 \subfloat[]{\includegraphics[width=.19\textwidth]{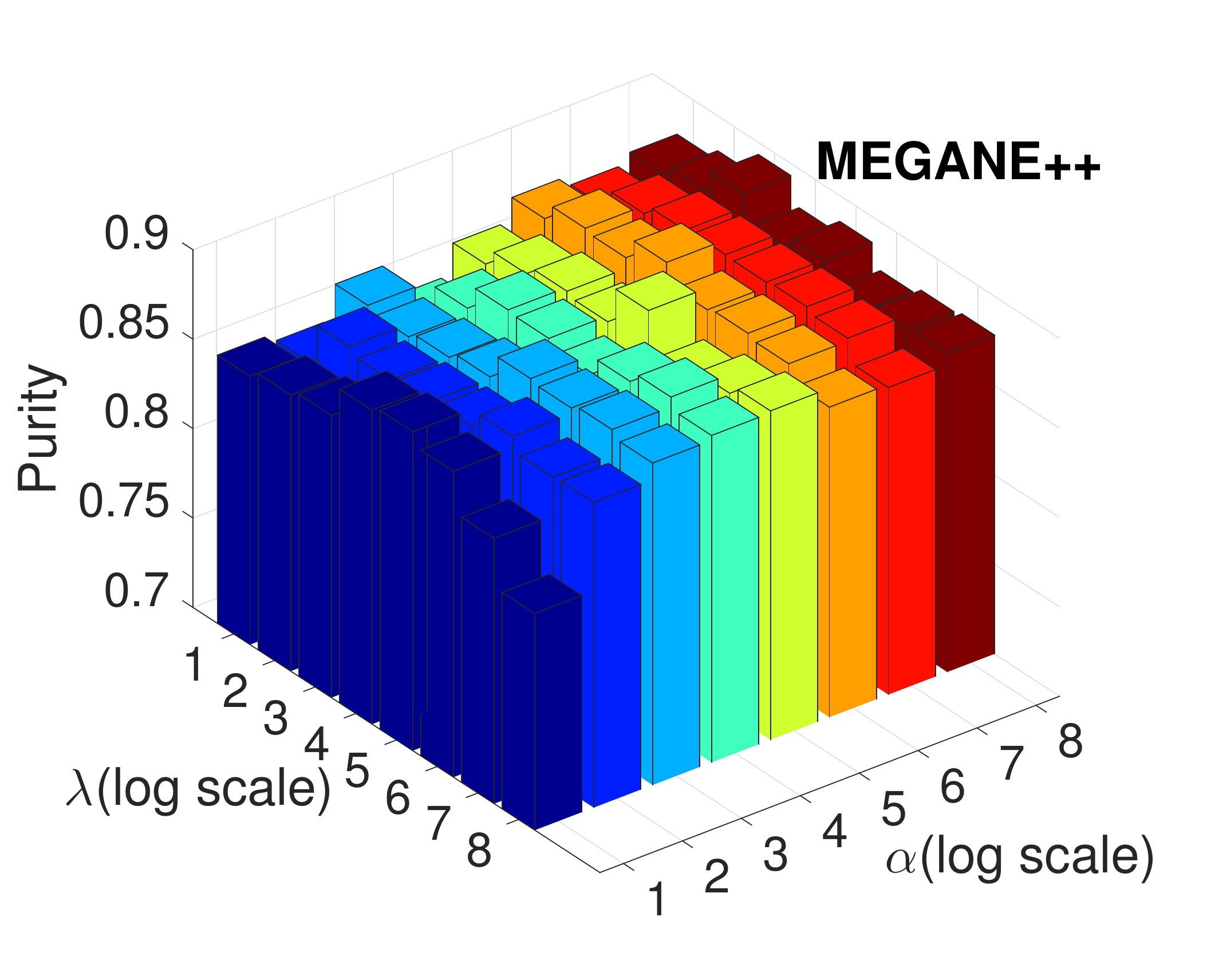}}
& \subfloat[]{\includegraphics[width=.19\textwidth]{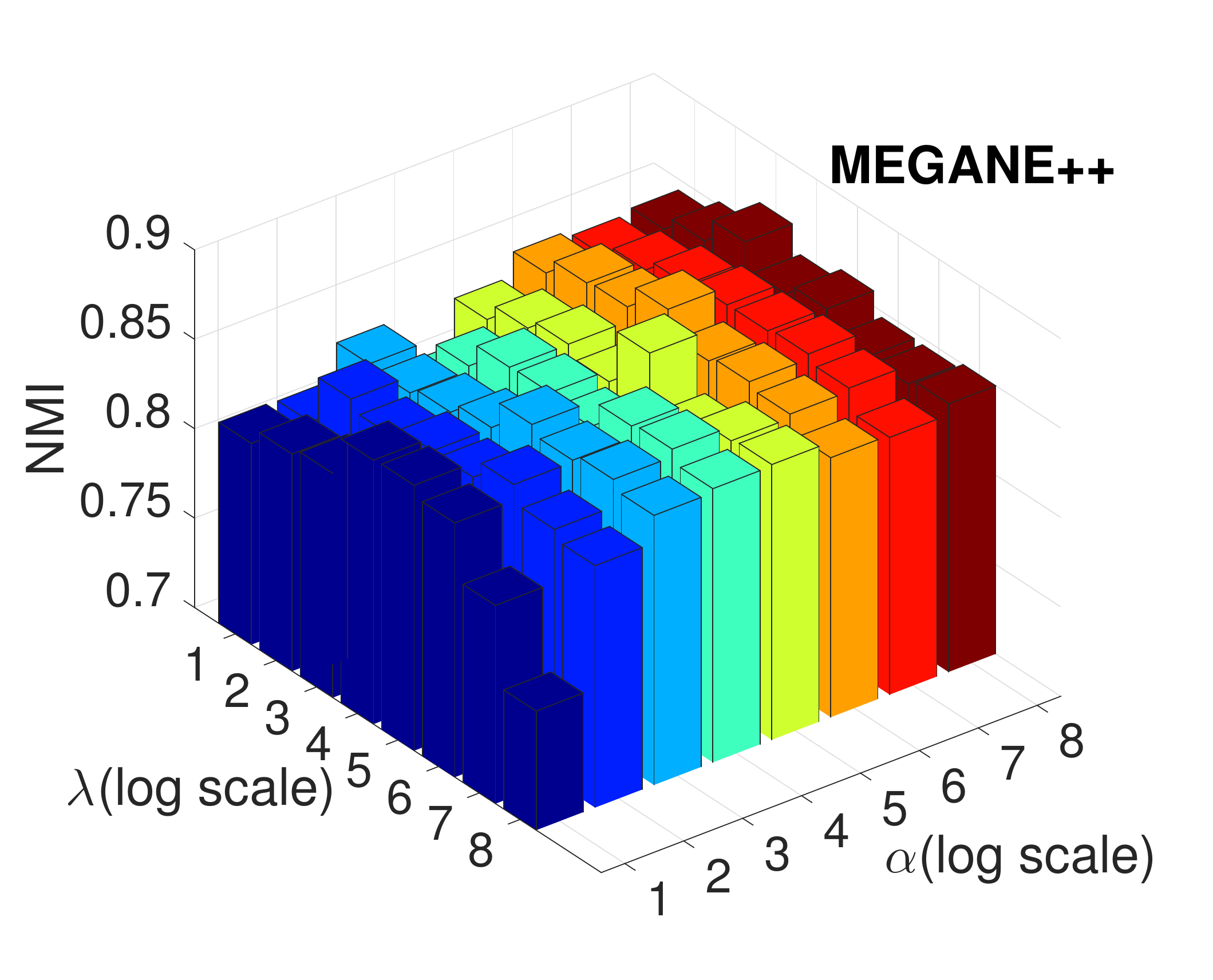}}
\end{tabular}
\caption{Parameters analysis: two metrics including Purity (a)(c)(e) and NMI (b)(d)(f). 
(a)-(b) analyzes $\lambda$ of {\sun}. 
(c)-(d) analyzes the embedding dimensions $R$ of
{\sun} and {\sunl}. 
(e)-(f) analyzes two parameters $\lambda$ and $\alpha$ of {\sunl}.}
\label{fig:parameters}
\end{figure}

\vspace{-2pt}
\subsection{Parameter Analysis}\label{sec:parameter}
In this section, we first analyze the parameter sensitivity of our methods as shown in Figure~\ref{fig:parameters}. 
We use two evaluation metrics, Normalized Mutual Information(NMI),
and Purity (both the larger, the better),
to evaluate the performances of our methods for clustering task. 
In Figure \ref{fig:parameters} (a)-(b), 
the penalty parameters $\lambda$ of {\sun} is used for minimizing the Frobenius Norm of embedding space $\mathbf{P}$ and $\mathbf{Q}$ in Eq.~(\ref{eq:obj_trans}),
and the best performance is achieved when $\lambda$ is set as 3.2768e-04.
From \ref{fig:parameters} (b)-(c), 
setting the embedding dimensions as 5 shows the best performance for both {\sun} and {\sunl}. 
{\sunl} outperforms {\sun} with the same number of embedding dimensions. 
The Figure \ref{fig:parameters} (e)-(f) show the two penalty parameters $\lambda$ and $\alpha$ of {\sunl}.
The penalty parameter $\lambda$ of {\sunl} is the same as that in {\sun}.
The penalty parameter $\alpha$ of {\sunl} is used for minimizing the Frobenius Norm of meta-graph similarity matrix $\mathbf{Y}$  and its embedding space in Eq.~(\ref{eq:obj}).
We find that $\lambda = 0.6711$ and $\alpha = 1.6$ produce the best performance of clustering task.

\begin{table}[t] \small
\centering
\caption{Time analysis: three tasks of {\sunl}: $R$ is the dimension of the embedding, and second is the time scale}
\label{table:time}
\begin{tabular}{|l|c|c|c|c|}
\hline
$R$      & 1 & 5  & 10  & 15\\ \hline 
DBLP (A2A)  & 0.338 & 4.693 & 9.731 &9.689           \\ \hline 
DBLP (V2V)  & 0.129 & 0.169 & 0.445  & 0.667         \\ \hline 
MOVIE (M2M) & 1.625 & 6.013 & 12.69 & 19.57         \\ \hline 

\end{tabular}
\end{table}
\vspace{-5pt}
\subsection{Time Analysis}\label{sec:time}

In this section, we evaluate the execution time of {\sunl}.
In Table \ref{table:time}, it shows the execution time is linear with respect to the embedding dimensions.
Based on the time complexity of {\sunl}, when we have a fixed size of dataset, the embedding dimensions $R$, and the number of views $N$ are linear with respect to the execution time.
Sometimes, {\sunl} can be early stopped when it is already converge, so as to the same time consuming of DBLP (A2A) with $R = 10$ and $R = 15$.
The same results are shown in the real testing on three tasks, which show the efficiency of {\sunl}.

\section{Related Work}\label{sec:relatedwork}





\subsection{Network Embedding}

Network embedding want to learn a low-dimensional representations from a
network. Previous traditional works~\cite{belkin2001laplacian} usually 
construct the affinity graph using the feature 
vectors of the vertexes and then compute the 
eigenvectors of the affinity graph. Some other groups
use matrix factorization to represent graph as adjacency matrix~\cite{ahmed2013distributed}.

Recently, DeepWalk~\cite{perozzi2014deepwalk} and LINE~\cite{tang2015line}
are proposed for learning the network embedding. 
Besides these two most popular node embedding methods,
many other network embedding are proposed recent years~\cite{chang2015heterogeneous,gui2016large,wang2016structural,chen2017task}.
\cite{chang2015heterogeneous,wang2016structural} learn the node embedding
by deep learning encoder methods.
However, none previous node embedding methods consider the meta-graph and its embedded 
meta-paths information.

\vspace{-5pt}
\subsection{Tensor Learning and Embedding}

Just like deep learning, tensor learning becomes very hot and popular topic in recent years due to the stronger computing capability and lower computation cost ~\cite{he2017kernelized,liu2015low,guo2014ga,lu2017multilinear,shao2015clustering,cao2017t,he2014dusk}. 
Coupled tensor matrix embedding tries to fuse multiple information sources where matrices and tensors sharing some common modes are jointly embedding \cite{ermics2015link}. 
A gradient-based optimization approach for joint tensor-matrix analysis is proposed
by Acar et al. \cite{acar2011all}. 

\subsection{Multi-view Learning}

Multi-view learning is a hot idea to think one object with different views \cite{sun2017sequential, sun2017contaminant,shao2016online,he2017multi}. In this paper, we think the HIN with different views such as meta-paths and meta-graph, and fuse the different information for node embedding.
However, none of these frameworks can be directly applicable 
to learn jointly embedding with a partial symmetric tensor and a symmetric matrix,
and also do not leverage meta-path and meta-structure information for similarity search in HINs.

\section{Conclusion and Future Work}\label{sec:conclusion}

In this paper, we proposed a new meta-graph-based relevance measure, {\em i.e.} GraphSim,
and two node embeddings, {\em i.e.} {\sun} and {\sunl}, by leveraging a meta-graph and its embedded meta-paths similarity information. 
In the experiment, {\sunl} shows better performance than other compared methods
in different tasks.
In the future, we can expend our proposed node embedding for a single meta-graph to
multiple meta-graphs node embedding in a HIN.
Meanwhile, we can utilize heterogeneous and homogeneous information together for node embedding.

\section{Acknowledge}
This work is supported in part by NSFC through grants No. 61503253 and 61672313, NSF through grants No. IIS-1526499, IIS-1763325, and CNS-1626432, and NSF of Guangdong Province through grant No. 2017A030313339.


\bibliographystyle{plain}
\bibliography{reference.bib}

\end{document}